\documentclass[10pt,conference]{IEEEtran}
\IEEEoverridecommandlockouts
\usepackage{cite}
\usepackage{amsmath,amssymb,amsfonts}
\usepackage{algorithmic}
\usepackage{graphicx}
\pdfoutput=1
\usepackage[dvipsnames]{xcolor}
\usepackage{colortbl}
\usepackage[utf8]{inputenc}
\usepackage[english]{babel}
\usepackage[T1]{fontenc}
\usepackage{hyperref}
\hypersetup{
  colorlinks   = true, 
  urlcolor     = blue, 
  linkcolor    = red, 
  citecolor   = green 
}
\usepackage{lipsum}

\usepackage{textcomp}
\usepackage{centernot}
\usepackage[braket, qm]{qcircuit}
\usepackage{braket}
\def\BibTeX{{\rm B\kern-.05em{\sc i\kern-.025em b}\kern-.08em
    T\kern-.1667em\lower.7ex\hbox{E}\kern-.125emX}}
\usepackage{multicol}
\usepackage{wrapfig}
\usepackage{tabularx}
\usepackage{tikz}
\usepackage{pgfplots} 
\usetikzlibrary{pgfplots.groupplots}
\usetikzlibrary{intersections}
\pgfplotsset{compat=1.16}
\pgfplotsset{scaled y ticks=false}
\usepgfplotslibrary{fillbetween}
\usepackage{mathtools}
\usepackage{listings}
\usepackage{url}
\usepackage{booktabs} 

\usepackage{array}
\newcommand{\PreserveBackslash}[1]{\let\temp=\\#1\let\\=\temp}
\newcolumntype{C}[1]{>{\PreserveBackslash\centering}p{#1}}
\usepackage{comment}
\usepackage{multirow}
\usepackage{flushend}
\usepackage{float}
\usepackage{cleveref}
\usepackage{algorithm2e}
\RestyleAlgo{ruled}

\definecolor{darkmagenta}{rgb}{0.55, 0.0, 0.55}
\usepackage{float}
\usepackage{afterpage}

\newif\ifdraft
\drafttrue
\ifdraft
\newcommand{\lpnote}[1]{ {\textcolor{blue} { ***Lilly: #1 }}}
\newcommand{\kwnote}[1]{ {\textcolor{green} { ***Karen: #1 }}}
\newcommand{\aanote}[1]{ {\textcolor{purple} { ***Abhishek: #1 }}}
\newcommand{\mhnote}[1]{ {\textcolor{orange} { ***Max: #1 }}}
\else
\newcommand{\note}[1]{}
\newcommand{\lpnote}[1]{}
\newcommand{\kwnote}[1]{}
\newcommand{\aanote}[1]{}
\newcommand{\mhnote}[1]{}
\fi

\begin{document}

\title{Effective Embedding of Integer Linear Inequalities for Variational Quantum Algorithms}
\author{
\IEEEauthorblockN{Maximilian Hess\IEEEauthorrefmark{1}\IEEEauthorrefmark{4}, Lilly Palackal\IEEEauthorrefmark{1}\IEEEauthorrefmark{4}, 
Abhishek Awasthi\IEEEauthorrefmark{2}\IEEEauthorrefmark{4}, Karen Wintersperger\IEEEauthorrefmark{3}\IEEEauthorrefmark{4}}
\IEEEauthorblockA{\IEEEauthorrefmark{1}Infineon Technologies AG, Munich, Germany}
\IEEEauthorblockA{\IEEEauthorrefmark{2}BASF Digital Solutions GmbH, Ludwigshafen am Rhein, Germany}
\IEEEauthorblockA{\IEEEauthorrefmark{3}Siemens AG, Munich, Germany}
\IEEEauthorblockA{\IEEEauthorrefmark{4}Quantum Technology and Application Consortium (QUTAC)}
\{maximilian.hess, lilly.palackal\}@infineon.com, \\ abhishek.awasthi@basf.com, karen.wintersperger@siemens.com
}

\maketitle
 \thispagestyle{plain}
 \pagestyle{plain}

\pagebreak

\begin{abstract}
In variational quantum algorithms, constraints are usually added to the problem objective via penalty terms. For linear inequality constraints, this procedure requires additional slack qubits. Those extra qubits tend to blow up the search space and complicate the parameter landscapes to be navigated by the classical optimizers. In this work, we explore approaches to model linear inequalities for quantum algorithms without these drawbacks. More concretely, our main suggestion is to omit the slack qubits completely and evaluate the inequality classically during parameter tuning. We test our methods on QAOA as well as on Trotterized adiabatic evolution, and present empirical results. As a benchmark problem, we consider different instances of the multi-knapsack problem. Our results show that removing the slack bits from the circuit Hamiltonian and considering them only for the expectation value yields better solution quality than the standard approach. The tests have been carried out using problem sizes up to 26 qubits. Our methods can in principle be applied to any problem with linear inequality constraints, and are suitable for variational as well as digitized versions of adiabatic quantum computing.
\end{abstract}


\section{Introduction}\label{sec: introduction}
\textbf{Motivation}: Combinatorial optimization has many applications in industry settings, ranging from logistics to production and more general resource allocation problems.
One of the best studied combinatorial optimization problems is the knapsack problem. In this work, we specifically focus on the 0/1 multi-knapsack problem, a  binary optimization problem with linear objective and linear inequality constraints. Several classical methods exist for solving the problem or finding approximate solutions~\cite{Kellerer2004,laabadi,wilbaut}. Like most problems in combinatorial optimization, the knapsack problem and its variants are NP-hard, which means that it is unlikely that a polynomial time algorithm to solve the problem exists. Especially for quantum computing, dealing with inequality constraints comes with issues like more qubits and harder convergence.


Recent advances in quantum computing have opened up new possibilities for solving such optimization problems. The most prominent approaches for the currently limited quantum hardware devices are quantum annealing and variational approaches for the circuit model~\cite{Cerezo_2021}. Both of these approaches require the optimization problem in the form of a Hamiltonian, whose ground state energy is sought utilizing the principles of quantum mechanics ~\cite{glover}. 
Variational quantum algorithms train a parameterized quantum circuit with classical optimizers and thus tend to handle errors better than other powerful algorithms such as Shor’s factoring algorithm~\cite{shor} or Grover’s search~\cite{grover}. As a consequence, they do not require fully error-corrected qubits, and therefore have the potential to be useful even with relatively few physical qubits. However, the power of such variational quantum algorithms is not fully understood yet. In order to gain more insights into the usefulness of such algorithms, implementations for solving various problems, such as combinatorial optimization or chemical simulation, have been derived. Several attempts at solving combinatorial optimization problems with the help of the Quantum Approximate Optimization Algorithm (QAOA) have yielded rather negative results~\cite{qutac_knapsack}. Therefore, a number of suggestions for improvement have been developed, such as warm-started QAOA \cite{Egger_2021}, R-QAOA \cite{Bravyi2022_RQAOA} or improvements of the classical parameter optimization. 

Integer linear programs with inequality constraints such as the knapsack problem tend to scale rather badly in the required number of qubits when following straight forward formulations as Quadratic Unconstrained Binary Optimization (QUBO) problems with extra variables (slack bits) as given in \cite{lucas_2014}. 
These slack bits are needed to convert inequality constraints into equality constraints for which one can formulate quadratic penalty terms. In the knapsack problem these inequalities (and thus the slack bits) originate in modeling the capacity constraint of a knapsack. The incorporation of slack bits in quantum algorithms comes with two main challenges. One is simply the requirement of  qubits to accurately model the problem on the quantum hardware, and second is the convergence of the slack bits. It has been shown in earlier works how these two issues impact the performance of quantum algorithms~\cite{quintero,nazarro,qutac_knapsack}.

One of the primary problems with the convergence of slack bits is related to the increased complexity of the optimization landscape. The addition of slack bits expands the solution space, making it more difficult for the algorithm to navigate and find the optimal solution. The presence of a larger solution space can result in a higher number of local optima, causing the optimization process to get stuck at sub-optimal solutions or to converge slowly. This issue becomes more prominent as the number of slack bits increases and as the problem size grows. Furthermore, the interaction between the slack bits and the main variables of the problem can create intricate dependencies, making it challenging to find a balance between satisfying constraints and optimizing the objective function. This can lead to difficulties in converging to a solution that simultaneously satisfies all constraints and optimizes the objective function. Additionally, the optimization process with slack bits requires careful tuning of various algorithmic settings, such as the assignment of values to the slack bits, the selection of classical optimization techniques, and the choice of termination criteria. Poorly chosen parameter values or optimization strategies can hinder the convergence of the algorithm, leading to sub-optimal solutions or premature convergence.

\textbf{Related work}: There have been a few studies made to address the issues of slack bits in some recent works. Koretsky~\emph{et al.} introduce a novel technique to embed inequalities into QAOA, without requiring extra qubits in the Hamiltonian formulation~\cite{koretsky}. They propose to incorporate real-valued slack variables which are optimized classically, while only the problem bits are embedded in the quantum circuit. The technique is helpful in avoiding the binary slack variables, but requires the QAOA ansatz to find out the ground state of a changing Hamiltonian, along with the classical optimization of QAOA angles and the continuous slack variables, at every iteration of the classical loop. The work of Braine~\emph{et al.} is an extension to the previous work of Koretsky~\emph{et al.}, proposing to initially optimize the binary and the continuous variables simultaneously, and then perform an iterative process of fixing the continuous variables and optimizing the binary variables with a quantum algorithm~\cite{braine}. This approach allows for a correct convergence of the continuous variables, depending on the binary values. However, the issue of a changing Hamiltonian still persists, and at different classical iterations, the VQE or QAOA ans\"atze are required to find ground state(s) of different Hamiltonians. There is also some work for the same kind of optimization problems (including inequalities) using quantum annealing. Recently, Montane-Barrera~\emph{et al.} proposed a technique to convert an inequality ($A\cdot\mathbf{x}\geq0$) to a QUBO without requiring slack bits, by approximating the exponential of the inequality terms ($e^{-A\cdot\mathbf{x}}$) up to quadratic terms using Taylor expansion~\cite{montanezbarrera}. The authors implement this approach for the TSP and knapsack problem, and propose an extension for QAOA. However, this approach requires the coefficients of the linear and quadratic terms to be pre-optimized in order to bring the optimal solution of the original combinatorial optimization problem as close as possible to the ground state of the cost Hamiltonian. Unfortunately, there is no clear strategy for finding suitable coefficients. Thus, for a general problem instance, it becomes difficult to find a Hamiltonian with the desired ground state.

\textbf{Key contributions}:
In this work, we propose hybrid quantum algorithms based on QAOA and Trotterized Adiabatic Evolution (TAE) which handle inequality constraints in the classical counterpart without introducing additional slack bits. This results in a lower number of required qubits, and a better convergence to the ground state compared to the standard QUBO formulation with slack bits.

Sec.~\ref{sec:methods} introduces the two quantum algorithms, QAOA and TAE as well as the core idea of evaluating inequalities classically in a no-slack-QUBO instead of the standard QUBO formulation for inequality constraints. In Sec.~\ref{sec: the multi knapsack problem} the multi-knapsack problem is introduced for which we numerically evaluate the presented approaches with experiments described in Sec.~\ref{sec: Experimental setup}. We discuss their results in Sec.~\ref{sec: results} and give a conclusion and outlook in Sec.~\ref{sec: conclusion and outlook}.

\section{Methods}\label{sec:methods}
\subsection{Quantum Approximate Optimization Algorithm (QAOA)}\label{sec:qaoa_algo}
QAOA is a popular variational algorithm devised to produce approximate solutions for combinatorial optimization problems~\cite{farhi,zhou}. Suppose $\mathcal{H}$ is an Ising Hamiltonian with integer coefficients, whose ground state corresponds to the optimal solution of a combinatorial optimization problem. The ingredients of QAOA comprise
\begin{itemize}
 \item the \textit{phase separation operator} $U(\mathcal{\mathcal{H}},\gamma) \coloneqq e^{-i\gamma \mathcal{\mathcal{H}}}$, where $\gamma \in [0,2\pi]$ is a free parameter and
 \item the \textit{mixing operator} $U(\mathcal{B},\beta)=e^{-i\beta \mathcal{B}}$ arising from the mixing Hamiltonian $\mathcal{B}=\sum_{j=1}^n \sigma_j^x$, where $\sigma_j^x$ denotes the tensor product of $\sigma_x$ acting on qubit $j$ with identity operators acting on all other qubits. Again, $\beta \in [0,\pi]$ is a free parameter.
\end{itemize}
The QAOA quantum circuit is applied to the ground state of the mixer Hamiltonian (Pauli-X) via alternating parameterized mixing and phase separation operators. Note that for a minimization problem we require either the initial state to be $\ket{-}^{\otimes n}$ or the mixer Hamiltonian to be $\mathcal{B}=-\sum_{j=1}^n \sigma_j^x$. In this work, we formulate the QUBO as a minimization problem and initialize QAOA with $\ket{-}^{\otimes n}$. One pair of mixing and phase separation operator is referred to as one layer. After the application of $p$ layers of parameterized operators, we arrive at the variational quantum state
\begin{equation}
\small
\begin{split}
 \ket{\psi_{\boldsymbol{\gamma}, \boldsymbol{\beta}}} = & \prod_{l = p}^{1} U(\mathcal{B},\beta_l) \cdot U(\mathcal{\mathcal{H}},\gamma_l)\cdot \ket{-}^{\otimes n} \;,
\end{split}
 \label{eq:qaoa_state}
\end{equation}
where $\ket{-}^{\otimes n}= (\ket{0}-\ket{1})^{\otimes n}/\sqrt{2^n}$ is a uniform superposition of $n$ qubits. The variational quantum state $\ket{\psi_{\boldsymbol{\gamma}, \boldsymbol{\beta}}}$ is parameterized by $2p$ parameters $\gamma_1,\dots, \gamma_p \in [0,2\pi]$ and $\beta_1, \dots, \beta_p \in [0,\pi]$. We then measure in the computational basis in order to compute the expectation value $F_p(\boldsymbol{\gamma}, \boldsymbol{\beta}) = \bra{\psi_{\boldsymbol{\gamma}, \boldsymbol{\beta}}}\mathcal{H}\ket{\psi_{\boldsymbol{\gamma}, \boldsymbol{\beta}}}$. 
The parameters $\boldsymbol{\gamma}, \boldsymbol{\beta}$ are varied in order to minimize $F_p$. Once the optimizer terminates at optimal or close to optimal parameters $\boldsymbol{\gamma}^*, \boldsymbol{\beta}^*$, we expect to sample an optimal or close to optimal solution, provided a reasonable number of measurement shots. 

\subsection{Trotterized Adiabatic Evolution}\label{sec: trotterized adiabatic evolution}
In this work we also test the trotterized version of the adiabatic evolution (TAE) for our problem Hamiltonians. In the literature, TAE is also sometimes referred to as digitized quantum annealing~\cite{Kovalsky_2023,kockum2024lecture,Hegade_2022}. We present a short formal description of TAE and draw an analogy with QAOA. 

Adiabatic quantum computation is inspired from the adiabatic theorem, which states that if a system is in the $n$th energy eigenstate of an initial Hamiltonian $\mathcal{H}_{\text{init}}$, then the system will remain in the $n$th energy eigenstate of the final Hamiltonian $\mathcal{H}$, provided that the Hamiltonian varies sufficiently slowly compared to the energy gaps between the $n$th eigenvalue and the remaining spectrum of the instantaneous Hamiltonian~\cite{kockum2024lecture}. This evolution of the Hamiltonian is governed by an annealing time $\mathcal{T}$ and an annealing schedule $s(t)$. The instantaneous Hamiltonian at any time $t$ ($0\leq t \leq \mathcal{T}$) is given by
\begin{equation}
\small
\mathcal{H}(t) = (1-s(t))\cdot \mathcal{H}_{\text{init}} + s(t)\cdot \mathcal{H}\;, 
\label{adb1}
\end{equation}
where $s(0)=0$ and $s(\mathcal{T})=1$. The initial Hamiltonian $\mathcal{H}_{\text{init}}$ is chosen such that its ground state can be easily prepared, and the most obvious and frequent choice is the Pauli-X operator.

The digitized evolution under the action of the Hamiltonian $\mathcal{H}(t)$ over the time $\mathcal{T}$ is given by $\hat U(\mathcal{T})$, where
\begin{equation}
\small
\hat U (\mathcal{T}) = \tau \exp{\left[-i \int_{0}^{\mathcal{T}} \mathcal{H}(t) \,dt \right]} \approx \prod_{l=p}^{l=1}\exp{\left[-i \mathcal{H}(l\delta t) \delta t \right]}\;, 
\label{adb2}
\end{equation}
where $p$ is some large integer also referred to as the number of Trotter steps and $\delta t =\mathcal{T}/p$. Note that we have reversed the indices in the product operator to account for the time-ordering operator $\tau$, \emph{i.e.}, the first trotter-step ($l$=1) acts first, followed by subsequent steps on an initial quantum state. Substituting the time dependent Hamiltonian from Eq.~\eqref{adb1}, we have
\begin{equation}
\small
\hat U (\mathcal{T}) \approx \prod_{l=p}^{l=1}\exp{\left[-i ((1-s(l\delta t)) \mathcal{H}_{\text{init}} + s(l\delta t) \mathcal{H}) \cdot \delta t \right]}\;.
\label{adb3}
\end{equation}
As the above digitization still requires big blocks of quantum gates, it is desirable to decompose the exponential operators into smaller blocks. From the Suzuki–Trotter expansion of the first order we know that for two non-commuting operators $\hat O_1$ and $\hat O_2$ and a sufficiently small $\delta t$, we can decompose the exponential operator as,
\begin{equation}
\small
\begin{split}
e^{i(\hat O_1+\hat O_2)\delta t}=e^{i\hat O_1\delta t} e^{i\hat O_2\delta t} + \mathcal{O}(\delta t^2) \;,
\label{trotter}
\end{split}
\end{equation}
where $\mathcal{O}(\delta t^2)$ is the Trotter error which vanishes for $\delta t\rightarrow 0$. Applying the above Trotterization to the time evolution operator we obtain
\begin{equation}
\small
\begin{split}
\hat U (\mathcal{T}) & \approx \prod_{l=p}^{l=1}\exp{\left[-i (1-s(l\delta t)) \mathcal{H}_{\text{init}} \delta t \right]} \exp{\left[-i s(l\delta t) \mathcal{H} \delta t \right]}\;.
\label{adb4}
\end{split}
\end{equation}
We can now use this as a digitized adiabatic evolution operator to approximate the ground state of the final Hamiltonian $\mathcal{H}$. If we take $\mathcal{H}_{\text{init}}=\sum_{j=1}^n \sigma_j^x$, and the initial state as $\ket{-}^{\otimes n}$, we can approximate the ground state $\ket{\psi^*}$ of $\mathcal{H}$ as $\ket{\psi^*} \approx \hat U (\mathcal{T})\ket{-}^{\otimes n}$. Using this understanding, we can now draw an analogy of the above trotterized adiabatic evolution with QAOA. In Eq.~\eqref{adb4}, if we substitute $(1-s(l\delta t)) \cdot \delta t \longrightarrow \beta_l$, and $s(l\delta t) \cdot \delta t \longrightarrow \gamma_l$, we can express final quantum state of QAOA as,
\begin{equation}
\small
\begin{split}
\ket{\psi_{\boldsymbol{\gamma}, \boldsymbol{\beta}}} = \prod_{l=p}^{l=1}\exp{\left[-i \beta_l \mathcal{H}_{\text{init}} \right]} \exp{\left[-i \gamma_l \mathcal{H} \right]} \cdot \ket{-}^{\otimes n}\;.
\label{qaoa_trotter}
\end{split}
\end{equation}
It can be seen that the above quantum state is identical to Eq.~\eqref{eq:qaoa_state}. The number of trotter steps ($p$) acts as the number of layers in QAOA. The crucial difference between TAE and QAOA is that TAE uses fixed time segments while QAOA relies on variational parameters which need to be optimized by classical means, necessitating many iterations of the quantum circuit~\cite{kockum2024lecture}. We explain the choice of the annealing schedule $s(t)$ and $\delta t$ in Sec.~\ref{sec:annealing_schedules}.


\subsection{Objective function and QUBO variants}\label{sec:qubo}
This work is motivated by better modelling and optimizing linear inequality constraints. In this section, we first present the conventional method to model inequalities by the introduction of slack bits into a QUBO. Next, we describe two alternative ways of dealing with the inequality constraints.
\subsubsection{Modelling of linear inequality constraints for quantum algorithms}~\label{sec:ineq_Sec_1}
There exist well known routines to convert linear equality and inequality constraints to QUBO form~\cite{glover}. The equality constraints of the type $a^T \mathbf x = b$ (where, $\mathbf x \in \{0,1\}^n, a \in \mathbb{R}^n, b \in \mathbb{R}$) are modeled as $P \cdot (a^T\mathbf  x - b)^2$ with $P\gg0$, such that the minimization of $P \cdot (a^T\mathbf  x - b)^2$ leads to a solution which satisfies $a^T\mathbf  x = b$. The linear inequality constraints of the type $c^T\mathbf x \leq d$ (where $c \in \mathbb{Z}^n, d \in \mathbb{Z}^+$) are first converted to an equality constraint by introducing a slack variable $r\in\mathbb{Z}^+$, such that $c^T\mathbf x + r = d$ where $r\leq d$. Since the exact value of the positive integer $r$ depends on $\mathbf x$, $r$ acts as a variable and is represented using binary slack bits as $r = \sum_{b=0}^{\lfloor \log d \rfloor} 2^b\cdot y_b$, where $y_b\in \{0,1\}$ $\forall b$. Substituting the value of $r$ to the newly formed equality constraint, we obtain $c^T\mathbf x + \sum_{b=0}^{\lfloor \log d \rfloor} 2^b\cdot y_b = d$. Note that the integer variable $r$ is discretized such that it ensures $r\leq d$. Using these slack bits, one can easily convert the original inequality constraint to a penalty QUBO term as 
\begin{equation}
\small
\text{penalty-qubo} = P\cdot \left(c^T\mathbf x + \sum_{b=0}^{\lfloor \log d \rfloor} 2^b\cdot y_b - d\right)^2 \;,
\label{ineq_qubo}
\end{equation}
where $P$ is again some large positive integer, also known as the penalty coefficient. Minimization of this QUBO requires an assignment of $\mathbf x$ and $r$ in such way that Eq.~\eqref{ineq_qubo} goes to zero, and thus satisfies the original inequality constraint $c^T\mathbf x \leq d$. Formally, the full QUBO for an optimization problem with such an approach can be modeled as follows. Suppose we are given an optimization problem
\begin{equation}
\small
\min_{\mathbf x\in \{0,1\}^n} f(\mathbf x) \;, \hspace{2em} \text{s.t. } c^T \mathbf x \leq d \;,
\label{eq:inequality_constraint}
\end{equation}
where $f:\{0,1\}^n \rightarrow \mathbb{R}$ is a quadratic objective function and $c \in \mathbb{Z}^n \text{ and } d \in \mathbb{Z}^+$ define a linear inequality constraint. The corresponding QUBO for the optimization problem can be written as $\mathcal{H} = f(\mathbf x) + P\cdot \left(c^T\mathbf x + \sum\nolimits_{b=0}^{\lfloor \log d \rfloor} 2^b\cdot y_b - d\right)^2$.
While the introduction of slack variables presents a conceptually valid method of incorporating inequality constraints into the objective function, the performance of most variational quantum protocols usually suffers significantly. On one hand, this is because in most protocols, one slack variable corresponds to one additional qubit. The number of qubits is already one of the limiting resources in most present-day quantum systems. With additional slack qubits, the range of problems for which tests can be run on quantum systems or quantum simulators is even more limited than it is without them. While, on the other hand, shallow-depth QAOA and VQE are heuristic methods which are probabilistic at their core. Additional slack qubits inflate the space from which we sample and it makes it much harder to find the "the needle in the haystack".

\subsubsection{Remove slack bits}~\label{sec:ineq_Sec_2}
The convergence of slack bits in quantum variational algorithms such as QAOA is a major issue and there have been some attempts in better incorporating the inequality constraints for quantum algorithms. A novel technique to embed the inequalities is proposed by Koretsky~\emph{et al.}~\cite{koretsky}. Given an inequality of the form $c^T \mathbf x\leq d$, the inequality gap $\boldsymbol\alpha=d-c^T \mathbf x$ is treated as a real-valued variable which is dependent on $\mathbf x$. The variable $\boldsymbol\alpha$ is plugged into the QUBO formulation as a variable and is optimized in the classical subroutine of the QAOA algorithm, along with the $\boldsymbol\beta$ and $\boldsymbol\gamma$ parameters, as explained in Sec.~\ref{sec:qaoa_algo}. Using this approach, one can model the optimization problem in Eq.~\eqref{eq:inequality_constraint} as a QUBO, given by
\begin{equation}
\small
\mathcal{H}(\boldsymbol\alpha ) = f(\mathbf x) + P\cdot\left(c^T\mathbf x+\boldsymbol\alpha-d\right)^2, \hspace{1em} \boldsymbol\alpha \geq 0\;,
\label{eq:alpha_hamiltonian}
\end{equation}
which needs to go to zero, for the original inequality to be satisfied. Although this approach helps us to avoid the slack bits, there are a few other issues which need to be considered. One of the issues is the number of $\boldsymbol\alpha$-values to optimize. The optimal solution of Eq.~\eqref{eq:alpha_hamiltonian} indeed consists of a single optimal value of $\boldsymbol\alpha$, whose value corresponds to the inequality gap $d-c^T\mathbf x$. However, quantum optimization algorithms such as QAOA or VQE, rely on sampling of several states via measurement, at each iteration. Since the values of $\boldsymbol\alpha$ depend on $\mathbf x$, each distinct sample of $\mathbf x$ requires a distinct corresponding $\boldsymbol\alpha$ value during the optimization process. Thus, the number of $\boldsymbol\alpha$ parameters to optimize turns out to be equal to the number of distinct $\mathbf x$ samples drawn at each iteration. Another drawback is that the Hamiltonian itself is changing during the optimization: After each classical iteration of QAOA, we are expecting the algorithm to approximate the ground state of a different Hamiltonian, since each iteration might provide different $\boldsymbol\alpha$ values.

Nonetheless, motivated from this work, we approach the inequality constraints in a similar fashion and try to address the two issues mentioned above.


\paragraph{Computing $\boldsymbol\alpha$ and the expectation value}\label{comp_alpha_exp}
We propose not to take $\boldsymbol\alpha$'s as variables for the optimization process, instead compute them on-the-fly for each sample, during each iteration. In most cases the expectation value for the QAOA algorithm is computed classically, taking all the samples into account at each iteration. To compute the expectation value of the sampled quantum state against the Hamiltonian, one can equivalently compute the objective function value of the resultant basis-states and weight them with the corresponding probabilities $p_s$. Thus, to compute the expectation value for each sample of the optimization problem in Eq.~\eqref{eq:inequality_constraint}, we compute $\alpha_s$ for each basis state $\mathbf x_s$ as $\alpha_s = \max\{0,d-c^T\mathbf x_s\}$, and compute the expectation value $\mathbb{\tilde E}$ as
\begin{equation}
\small
\begin{split}
\mathbb{\tilde E} &= \sum\limits_{s=1}^{S} p_s\cdot \left[f(\mathbf x_s) + P\cdot\left(c^T\mathbf x_s +\alpha_s - d\right)^2 \right] \;, \\
&= \sum\limits_{s=1}^{S} p_s\cdot \left[f(\mathbf x_s) + P\cdot\left(\max\{0,c^T\mathbf x_s-d\}\right)^2 \right] \;,
\end{split}
\label{alpha_ineq}
\end{equation}
where $P$ is some large positive penalty coefficient. This approach corresponds to a direct evaluation of the inequality constraint: The penalty term in Eq.~\eqref{alpha_ineq} ensures that only the infeasible bitstrings which do not satisfy $c^T\mathbf x_s\leq d$ are penalized proportional to the excess $(c^T\mathbf x_s- d)^2$. The bitstrings which satisfy $c^T\mathbf x_s\leq d$ are assigned zero penalty. This is in contrast to the regular encoding using slack bits as shown in Eq.~\eqref{ineq_qubo}, where even the $x$-bitstrings which satisfy the inequality constraint can be assigned some penalty due to non-convergence of the corresponding slack bits. Note that we are relying on the fact that this way of computing the expectation value is done classically. 
\paragraph{Problem Hamiltonian formulation}\label{problem_hamiltonian}
The QAOA circuit requires a problem Hamiltonian to define the parameterized phase separation operator. For optimization problems with inequalities of the type $c^T\mathbf x\leq d$, we propose 
\begin{equation}
\small
\mathcal{\tilde H} = f(\mathbf x) + P\cdot (c^T \mathbf x - d)^2 \;,
\label{alpha_hamiltonian}
\end{equation}
\emph{i.e.}, the inequality is treated as an equality in the circuit. Although the ground state of this Hamiltonian does not necessarily satisfy the inequality constraint, the evaluation of the $x$-bitstrings and minimization of the expectation value in Eq.~\eqref{alpha_ineq} can guide the QAOA to the approximate ground state of the original problem by optimizing the circuit parameters. This concept can easily be applied to higher dimensional inequality constraints as well as problem settings involving other constraints, which can be encoded without requiring slack bits. The authors of this manuscript are well aware that this ansatz is not supposed to yield the ground state of the problem for $p \rightarrow \infty$, since a different Hamiltonian is used in the QAOA phase separation operator. However, in practise, a finite number of circuit layers is used and the ground state is usually only approximated by a final superposition state which contains the ground state with a sufficiently high probability that ensures it being sampled at least once. In Sec.~\ref{sec: results} we show with help of experiments, that this approach can work very well for the QAOA algorithm.

\subsubsection{Evaluate only logical bits}\label{sec:ineq_Sec_3}
In this section, we present another approach to deal with inequality constraints, which differs form the standard QAOA only in the expectation value evaluation. The proposal for this approach is to retain the original QUBO for the inequality with slack bits, as mentioned in Sec.~\ref{sec:qubo}. In doing so, we ensure that the ground state of the underlying problem Hamiltonian for the phase separation operator in QAOA corresponds to the optimal solution of Eq.~\eqref{eq:inequality_constraint}. However, for the computation of the final expectation value, we use Eq.~\eqref{alpha_ineq}, which ensures that we only penalize the logical bits ($x$-bitstrings) which violate the inequality $c^T\mathbf x\leq d$. Logical bits which satisfy $c^T\mathbf x\leq d$ regardless of incorrect $y$-bitstrings, are not penalized, which is desirable. Once the parameters of the QAOA algorithm are optimized, the objective function used for evaluating the resultant bitstrings are given by
\begin{equation}\small
\mathcal{F}(\mathbf x) = f(\mathbf x) + P\cdot\left(\max\{0,c^T\mathbf x-d\}\right)^2.
\label{eval_obj}
\end{equation}
\begin{figure}[ht]
 \centering
 \includegraphics[width=0.45\textwidth]{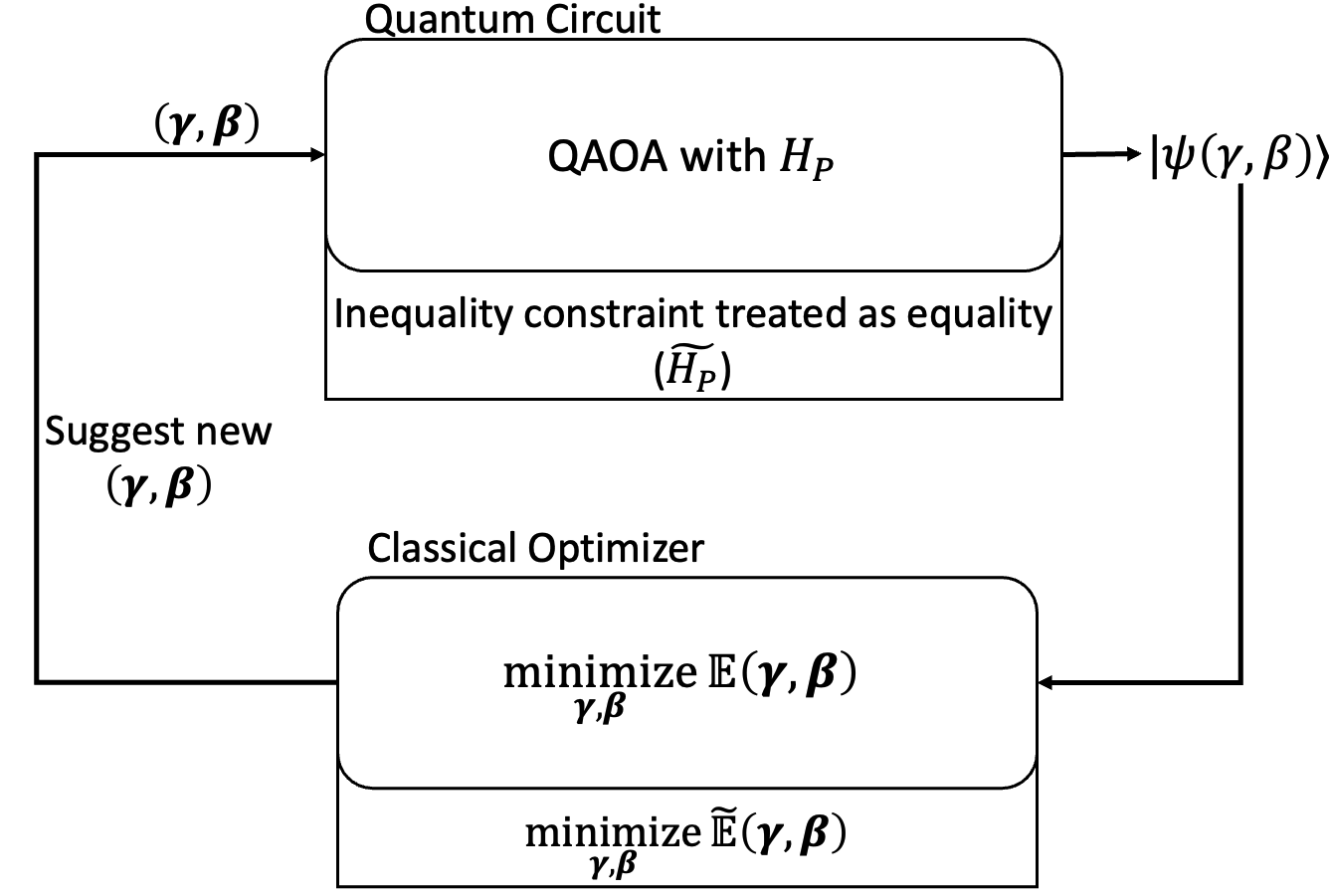}
 \caption{Algorithmic protocols discussed in Sec.~\ref{sec:qubo}. The notations used for the expectation values and the Hamiltonians are explained in Sec.~\ref{sec:qubo}.}
 \label{fig:alpha-QAOA_protocol}
\end{figure}
\subsection{Annealing schedule and parameter initialization}\label{sec:annealing_schedules}
The trotterized adiabatic evolution requires the definition of an annealing schedule $s(t)$, the total annealing time $\mathcal{T}$ and the number of trotter steps $p$ (or, in other words $\delta t$, which is equal to $\mathcal{T}/p$), as shown in Eq.~\eqref{adb3}. Given a fixed annealing time $\mathcal{T}$, the annealing schedule needs to be a function in $[0,\mathcal{T}]$ while satisfying $s(t=0)=0$ and $s(t=\mathcal{T})=1$. In this work, we utilize the sinusoidal schedule function $s(t)$ given by
\begin{equation}
\small
s(t) = \sin^2\left[ \frac{\pi}{2} \sin^2\left(\frac{\pi t}{2\mathcal{T}}\right)\right] \;,
\label{annealing_schedule}
\end{equation}
The more usual choice of the annealing schedule is a linear function $t/\mathcal{T}$, however, with our tests we found that the sinusoidal function worked better. Depending on the number of trotter steps $p$ one can set up the complete quantum circuit for TAE. The schedule $s(l\delta t)$ in Eq.~\eqref{adb4} can be substituted as
\begin{equation}
\small
s(l\delta t) = \sin^2\left[ \frac{\pi}{2} \sin^2\left(\frac{\pi l}{2 p}\right)\right], \text{ since } \delta t = \mathcal{T}/p \;.
\label{annealing_schedule1}
\end{equation}
Hence, for a given trotter step $l$ we can express the annealing schedule as a function of total trotter steps $p$. However, Eq.~\eqref{adb4} requires us to provide $\delta t$, the trotterization step size, as a constant input parameter. With the help of experiments, we estimate $\delta t=1$ and $\delta t=0.75$ to be good candidate values. Moreover, the value of $\delta t$ has been worked out also in the work of Sack and Serbyn~\cite{Sack2021quantumannealing}, where they show that $\delta t=0.75$ works well. Hence, using the sinusoidal schedule and the value of $\delta t$ we implement the TAE algorithm, given the total number of trotter steps $p$.

As for the QAOA initial parameters, note that we draw the analogy between the TAE and QAOA parameters $\boldsymbol\beta, \boldsymbol\gamma$ in Sec.~\ref{sec: trotterized adiabatic evolution}. Thus, we can utilize the discussed annealing schedule and $\delta t$ to find good initial parameters for QAOA. Formally, for any layer $l$ of QAOA we initialize the parameters as $\beta_l=(1-s(l\delta t))\cdot \delta t$ and $\gamma_l=s(l\delta t) \cdot \delta t$. Our results show that this initialization of initial parameters for QAOA works much better than random initialization of $\boldsymbol\beta$ and $\boldsymbol\gamma$.


\subsection{Classical optimizers}~\label{sec:classical_optimizer}
To perform the classical parameter optimization in QAOA, we utilize the gradient-based optimizer Adam~\cite{kingma2017adam}. Adam combines the Adaptive Gradient Algorithm (AdaGrad), that features an adaptive learning rate, and Root Mean Square Propagation (RMSProp), where the moving average of the gradient is used to normalize it. Our implementation of Adam is based on~\cite{Gray2024}. We also tested various other optimizers, such as COBYLA and the BFGS algorithm, for which we used the predefined implementations from Scipy~\cite{scipy_2022}. These optimizers performed well only with full state-vector simulation, however sampling via finite measurements produced premature termination. Using a finite number of samples always introduces sampling noise, thus, evaluating the circuit with the same parameters leads to different results. Among optimizers that can explicitly deal with noisy data, such as Stochastic Gradient Descent (SGD) and RMSProp, Adam performed best in our setting for all objective function variants and scenarios considered in this work.
\begin{figure}[htbp]
\centering
\includegraphics[width=0.45\textwidth]{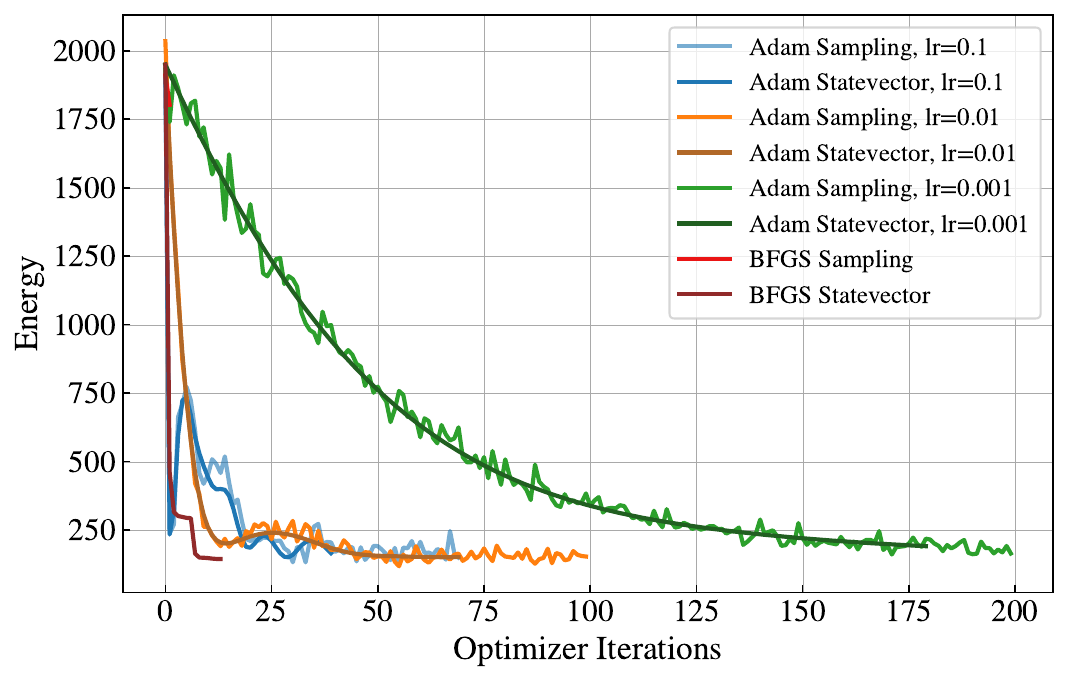}
\caption{Objective function (Energy) during optimization for a knapsack instance with 8 items and a single knapsack, without slack bits (evaluating $\tilde{H}_p$) and $p=3$ layers using the Adam and BFGS optimizers. The blue, orange and green curves denote the Adam optimizer with different learning rates. The light (dark) colors indicate sampling (full state-vector simulation). A single-penalty factor of $50$ and a sine schedule with $\delta t=0.75$ have been used (see Secs.~\ref{sec:annealing_schedules} and~\ref{sec:res_penalty_terms}).}
\label{fig:compare_optimizers}
\end{figure}
To apply Adam also to the optimization without slack bits, where the objective function is not differentiable (see Eq.~\eqref{alpha_ineq}), the gradient is evaluated numerically using finite differences. To save computation time, we added a stopping criterion to the optimizer to stop evaluation when a desired convergence in terms of a threshold is reached by the optimizer. Specifically, the optimization process is terminated when the following two conditions are fulfilled: The change in the moving average over the last $\omega$ function values is less than a threshold $f_\omega$, and the second derivative is positive and larger than a threshold $f_{sd}$ along all parameter dimensions. The latter avoids stopping at a plateau where the first and second derivative vanish. The two conditions are checked every $\omega$ steps and the moving average is calculated among the last $\omega$ functions values, including the current one at step $i$: $[f(\vec{x_i}), f(\vec{x}_{i-\omega})]$. With the help of several tests, we concluded that $\omega = 10$, $f_\omega = 10$, and $f_{sd} = 10$, workd best. The second derivative is also evaluated using finite differences. The step size for numerical calculation of both derivatives is set to $\varepsilon = 0.1$ in all cases. 

The optimal learning rate for the Adam optimizer was found by investigating the convergence behavior, and we plot the convergence of different optimizers with discussed parameters for a knapsack problem instance with 8 items and a single knapsack, using a QUBO formulation without slack bits, and $p=3$ circuit layers, in Fig.~\ref{fig:compare_optimizers}. The blue, orange and green curves show the expectation value of the QAOA objective function (energy) during optimization for different learning rates of the Adam optimizer, where the light colors denote the values for sampling and the dark colors indicate the results using full state-vector simulation. Overall, a learning rate of $l=0.01$ shows the best results, converging faster than $l=0.001$, and without the instabilities as observed for $l=0.1$. This behavior was observed consistently for all problem instances tested in this work, and thus, $l=0.01$ was used in all experiments shown here. Using the Adam optimizer with the settings described here, we also see a decrease in the final energy with the number of circuit layers for all algorithms and scenarios, indicating that indeed the approximation of the ground state improves as expected. 

Fig.~\ref{fig:compare_optimizers} also illustrates the effect of sampling noise on the performance and convergence of these optimizers. The results for the BFGS optimizer are plotted in red, which works well without sampling noise. This phenomenon of stability with full-state vector simulation was also observed with other classical optimizers we considered. However, in the presence of sampling noise, BFGS stops after only a very few iterations without having reached a minimum, as shown by the short light red curve. Considering experiments on real quantum devices, suffering also from various other sources of hardware noise on top of the sampling noise, the choice of appropriate classical optimizers for variational quantum algorithms becomes even more important. Building on our promising results for data with sampling noise, we would propose to use the Adam optimizer also for experiments on hardware.
\section{The Multi Knapsack Problem} \label{sec: the multi knapsack problem}
In this section, we give a short overview of the multi-knapsack problem which we use to test our methods.
\subsection{QUBO Formulation}
In the multi-knapsack problem, the task is to distribute $N \in \mathbb{N}$ items to $M \in \mathbb{N}$ knapsacks with capacities $c_j \in \mathbb{N}$. Every item $i \in \{0,...,N-1\}$ has a weight $w_i \in \mathbb{N}$ and a knapsack-dependent value $v_{i,j} > 0$, where $j \in \{0,...,M-1\}$. The objective of the problem is to maximize the value in the knapsacks without overstepping their respective capacities. In order to formulate the problem as a QUBO problem, we introduce binary decision variables $x_{i,j} \in \{0,1\}$ $\forall i,j$, such that $x_{i,j}=1$ indicates that item $i$ is assigned to knapsack $j$. 

Using this formalism we can now define the QUBO terms for all the constraints and the objective term. The summands of our QUBO formulation are explained below.
\begin{itemize}
\item[1.] The objective term is formulated such that our original maximization objective function is converted to a minimization problem, as shown below.
\begin{equation}
\small
H_{\text{obj}}= - \sum_{\forall i} \sum_{\forall j} v_{i,j} \cdot x_{i,j} \;.  
\label{eq_5}
\end{equation}
\item [2.] For the assignment constraint, we do not need to apply the usual procedure for inequalities, as for each item, there are only two scenarios which cover all feasible cases, namely the item being assigned to one knapsack or to no knapsack at all, so the quadratic term
\begin{equation}
\small
H_{\text{single}}= \sum_{\forall j} \left(\sum_{\forall i} x_{i,j}\right)\cdot \left(\sum_{\forall i} x_{i,j}-1 \right) \;.    
\label{eq_2}
\end{equation}
serves as a penalty term which is positive if and only if the assignment constraint is violated.
\item [3a.] The capacity $c_i$ of each knapsack $i \in \{0,\dots,M-1\}$ should not be exceeded. This may be achieved by introducing slack bits $y_{i,b}$ with binary expansion, as discussed above.
\begin{equation}
\small
\begin{split}
H_{\text{capacity}}=&\sum_{\forall i}\left[\left(\sum_{\forall j} w_j\cdot x_{i,j}\right)\hspace{-0.3em}+\hspace{-0.3em}\left(\sum_{b=0}^{\lfloor\log_2{c_i}\rfloor} \hspace{-0.3em}2^{b}\cdot y_{i,b}\right) \hspace{-0.2em}- \hspace{-0.2em}c_i \right]^2\;.
\end{split}
\label{eq_3}
\end{equation}
\item [3b.] As explained in Sec.~\ref{sec:ineq_Sec_2}, we can alternatively treat the inequality constraint as an equality and evaluate the inequality classically.
\begin{equation}
\small
   {\tilde{H}_{\text{capacity}}}= \sum_{\forall i}\left[\left(\sum_{\forall j} w_j\cdot x_{i,j}\right) - c_i \right]^2 . 
\end{equation}
\end{itemize}
With the above terms, we can now formulate the complete QUBO for multi-knapsack optimization problems in two variants.
\begin{itemize}
    \item \textbf{QUBO}: 
    \begin{equation}
    \small
    H_p = A\cdot H_{\text{single}} + B\cdot H_{\text{capacity}} + C \cdot H_{\text{obj}} \:.
    \label{eq_qaoa}
    \end{equation}
    \item \textbf{no-slack-QUBO}: 
    \begin{equation}
    \small
    \tilde{H}_p = A\cdot H_{\text{single}}  +B \cdot \tilde{H}_{\text{capacity}} + C \cdot H_{\text{obj}}\;.
    \label{eq_alpha_qaoa}
    \end{equation}
\end{itemize}

The coefficients $A,B > 0$ are the penalty weights, and $C>0$ is the objective weight. The minimization of $H_p$ results in an optimal solution to the multi-knapsack problem.

\subsection{Normalization}
The penalty coefficients and objective weights of a QUBO play an important role in deciding the full energy landscape of the optimization problem. In turn, this influences the performance of classical optimizers finding good QAOA parameters. For the knapsack problem in particular, the QUBO terms contain the item weights and values which in general leads to larger absolute numbers than e.g., for the MaxCut problem and accordingly to much more complicated energy landscapes. In this work, we normalize the coefficients of the Ising model to improve the energy landscape for all algorithms including TAE. Given an Ising model $\mathcal{H} = \sum_i h_i \sigma^x_i + \sum_{ij} J_{ij}\sigma^z_i\sigma^z_j$, we normalize $\mathcal{H}$ with the maximum absolute value of all the coefficients $\nu_{max} = \max\{ \underset{\substack{{\forall i}}}\max |h_i|,  \underset{\substack{{\forall {ij}}}}\max |J_{ij}| \}$. Normalizing the Hamiltonian can simplify the task of the classical optimizer. In fact, our experiments show better results with a normalization factor. However, in general, with normalization we possibly lose the guarantee that the optimal parameters lie in $[0,2\pi ]$.
\section{Experimental setup}\label{sec: Experimental setup}
\subsection{Setup}
In order to obtain a fair comparison of the different strategies for dealing with inequality constraints,~\emph{i.e.}, QUBO vs. no-slack-QUBO, we run simulations with the following settings. 
In total, 20 (multi) knapsack instances are considered on a range from 2 qubits (logical qubits in scenario 0) up to 26 qubits (logical and slack qubits in scenario 19). 
We run QAOA and the TAE as described in Sec.~\ref{sec: trotterized adiabatic evolution}, with the Hamiltonian being normalized and built either from the standard QUBO (Eq.~\eqref{eq_qaoa}) or the no-slack-QUBO (Eq~\eqref{eq_alpha_qaoa}). All computations are made by sampling the the quantum states via simulations, where the number of samples drawn is equal to $500$ times the number of qubits.
For evaluating an obtained state during the optimization in QAOA, we choose the expectation value corresponding to the respective problem Hamiltonian. 
In the final evaluation, we once include and once neglect the convergence of the slack bits when running the algorithms with slack bits, as explained in further detail in Sec.~\ref{sec:ineq_Sec_3}. The parameter optimization in QAOA is performed with the classical optimizer Adam with a learning rate of 0.01 as described in Sec.~\ref{sec:classical_optimizer}. For TAE, we run a single pass,~\emph{i.e.}, no optimization of classical parameters is required. After performing several tests with different annealing schedules, we have chosen the best performing setting which is the sinusoidal schedule with $\delta t = 0.75$. For QAOA, we derive the corresponding angles $\boldsymbol{\beta}, \boldsymbol{\gamma}$ from this schedule and use them as initial parameters.
Each setting is repeated 10 times and report the average performance in Sec.~\ref{sec: results}.

\subsection{Performance metrics}~\label{sec:metrics}
In the following we define the performance metrics used in our experiments. In practice, the final expectation value is not a KPI a user in interested in, rather in higher chances (probability) to receive good solutions to their problem. In general, the solution to the optimization problem is obtained after sampling from the final circuit as the bitstring with the lowest energy. When benchmarking algorithms on small problems with known optimal solutions, as in this work, the best solution can be compared with the optimum, and, e.g., an approximation ratio could be defined. As discussed in Sec.~\ref{sec:res_algos}, in our experiments, the known optimum was found in nearly all cases. Thus, to quantify the performance, we use the total probability to sample optimal bitstrings (in cases there are several optimal solutions), denoted by $P_{\text{opt}}$, and the total probability to sample any solution which is $90\%$ optimal, $P_{90}$. The latter includes all solutions which are valid (all penalty terms are equal to zero) and with $-H_{\text{obj}} \geq 0.9\cdot V_{\text{opt}}$, where $V_{\text{opt}}$ denotes the total value of the optimal item distribution. These quantities are derived as relative frequencies from the final distribution of sampled bitstrings:
\begin{equation}
\small
    P_{\text{opt}}\coloneqq \frac{\textrm{\#optimal solutions}}{\textrm{\#shots}}, \quad P_{90}\coloneqq \frac{\textrm{\#90\%-optimal solutions}}{\textrm{\#shots}}.
\label{eq:metrics}
\end{equation}
\section{Results}\label{sec: results}
\subsection{Comparison of algorithms and objective functions}~\label{sec:res_algos}
The different variants of objective functions and algorithms described in the previous sections are compared in terms of their performance. 
\begin{figure*}[htbp]
\centering
\includegraphics[width=1\textwidth]{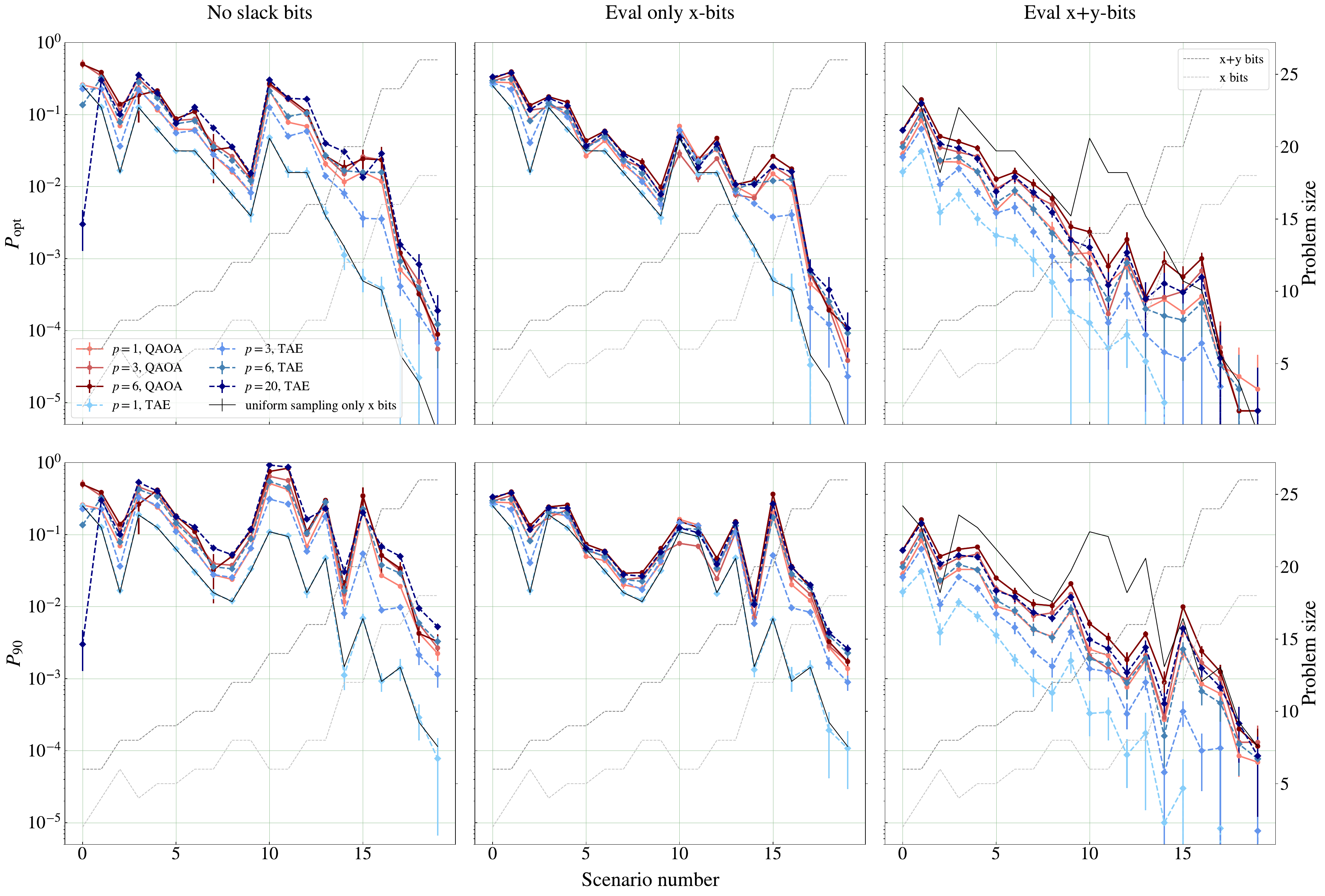}
\caption{Total probability of sampling optimal solutions $P_{\mathrm{opt}}$ (upper panels) and $90\%$ optimal solutions $P_{90}$ (lower panels) vs. scenario number. The columns indicate different objective functions, from left to right: Discarding slack bits at all, using slack bits, but evaluating only x-bits, and evaluating x- and y-bits, where we denote the logical bits with x and the slack bits with y. The results for QAOA (red circles) and TAE (blue diamonds) for various numbers of circuit layers are shown. We use a sinusoidal annealing schedule with $\delta t = 0.75$ for TAE and as initial parameters for QAOA. As a baseline, $P_{\mathrm{opt}}$ and $P_{90}$ for randomly drawing x-bitstrings are plotted as solid black lines. The dashed gray lines denote the problem sizes when counting the number of x-bits or x+y-bits. Missing points are zero and thus not displayed in the logarithmic scale.}
\label{fig:compare_algos}
\end{figure*}
In Fig.~\ref{fig:compare_algos}, the probability of sampling the optimal solution, $P_{\mathrm{opt}}$, and the probability of sampling a solution which is at most $90\%$ optimal, $P_{90}$ are plotted against the scenario number in the upper and lower panels, respectively. The three columns correspond to the different types of objective functions,~\emph{i.e.}, using no slack-bits, using slack-bits but evaluating only the x-bits and evaluating both x- \& y-bits.
The right $y$-axes shows the problem size in terms of only x-bits and x+y-bits (dashed gray lines). The latter problem size increases monotonically with the scenario number, while the number of x-bits sometimes decreases, e.g, for scenario 3 and 10. The probabilities are derived as described in Sec.~\ref{sec:metrics}. As a minimum baseline the values of $P_{\mathrm{opt}}$ and $P_{90}$ derived from uniform sampling are shown (solid black line), where 
\begin{equation}
\small
P_{\mathrm{opt}} = \frac{\textrm{\#optimal solutions}}{2^N}, \quad P_{90} = \frac{\textrm{\#90\%-optimal solutions}}{2^N}\;,
\end{equation}
with $N$ being the number of x-bits. These values correspond to randomly picking a configuration of items and knapsacks, which is why we only use the number of x-bits here. The red circles in Fig.~\ref{fig:compare_algos} denote the probabilities for QAOA, while the probabilities for TAE are shown as blue diamonds. Darker colors indicate a higher number of layers in the circuit. Note that for TAE, circuits with up to 20 layers have been run. We show results only for a selected number of layers for visual clarity. Overall, the performance in terms of $P_{\mathrm{opt}}$ and $P_{90}$ monotonically increases with the number of circuit layers in nearly all cases. The data points for TAE on scenario 0 without slack bits (left panels) are an exception, which we discuss in more detail below. The optimal probabilities decrease with the problem size: when evaluating x-bits (left and middle panels), the relevant quantity is the number of x-bits (light gray line), while for the data in the right panels, the problem size is given as x+y bits (dark gray line). This is expected, as the Hilbert space size increases, which can also be seen from the uniform sampling baseline. 

The circuits with no slack bits and with evaluation of x-bits yield values clearly exceeding the random baseline for all number of layers with optimized parameters (QAOA) and with at least two layers for TAE. In the case when the slack bits are also evaluated, the probabilities are much lower and lie below the baseline in mostly all cases. This reflects the fact, that a circuit with slack bits can produce optimal solutions in terms of x-bits, however with non-converged y-bits. Moreover, the overall state space size is much larger, when the y-bits are also taken into account, leading to smaller probabilities. For QAOA, we can see from Fig.~\ref{fig:compare_algos}, that discarding the slack bits completely, leads to comparable or even better results, even when the slack bits are not evaluated. Discarding the parameter optimization in TAE also does not decrease performance when the number of layers is increased accordingly, for instance, QAOA with $p=3$ and TAE with $p=6$ yield nearly equal probabilities. This holds especially when the slack bits are used in the circuit, but not evaluated. However, when the slack bits are discarded, the TAE approach can lead to worse results in some cases, as for scenario 0. This particular scenario consists of a single knapsack with capacity $c=10$, two items with weights $w = [4,6]$ and values $v=[19,16]$. The optimal solution packs the first item into the knapsack, corresponding to an excess capacity of 6. The Hamiltonian $\tilde{H_p}$ without slack bits treats the capacity constraint as an equality and thus over- and under-filling the knapsacks are penalized equally. This leads to an invalid solution, in which both items are packed with an excess capacity of 1, being preferred over the optimum. In TAE, the ground state of $\tilde{H_p}$ is approximated, which leads to the wrong solution being sampled with higher probability when the approximation is improved by adding more layers. This can be clearly seen from the blue data points for scenario 0 in the left panels of Fig.~\ref{fig:compare_algos}. 

Whether the ground states of $\tilde{H_p}$ and ${H_p}$ coincide depends on the specific problem under consideration. In general, TAE should be used with the full Hamiltonian ${H_p}$ including slack bits in the circuit, but only evaluating x-bits. As shown in the middle panel in Fig.~\ref{fig:compare_algos}, this leads to satisfying results, clearly adding benefit over standard QAOA.
This option needs more qubits, but TAE is much more efficient than QAOA in terms of runtime, as the circuit is only executed once. 
As a second option, the slack bits can be discarded when optimizing the angles in QAOA,~\emph{i.e.}, in the objective function (the expectation value computation), the inequality is being evaluated correctly and thus, the optimal solution will be preferred, as demonstrated by the QAOA results in the left panels of Fig.~\ref{fig:compare_algos}. Moreover, using the values of TAE as a starting point for the parameter optimization already improves performance compared to initializing the angles with random values for QAOA.



\subsection{Influence of penalty terms}~\label{sec:res_penalty_terms}
We now describe our choices for the penalty terms and the objective weights in Eqs.~\eqref{eq_qaoa}. The penalty QUBO terms ($H_{\text{single}}$ and $\text{H}_{\text{capacity}}$) are defined such they are always greater than or equal to zero, and their values are zero \textit{iff} the corresponding constraints are satisfied. To avoid optimizing the objective term at the expense of the penalty terms, we need to assign higher values for $A$ and $B$ than $C$. At the same time, we need to ensure that the penalty coefficients are not too high to avoid that the complete energy landscape of the optimization problem is dependent \textit{only} on the penalty terms of the QUBO. Hence, we need to find the penalty coefficients such that in all cases of an infeasible solution the sum of the penalty terms in the QUBO must be greater than the $H_{\text{obj}}$ value (which is always non-positive). In this work we take $C=1$ for all the algorithms, and take $A$ and $B$ equal to sum of all weights and sum of all values, formally
\begin{equation}
\small
A=B=\sum\nolimits_{j=0}^{N-1} w_j + \sum\nolimits_{i=0}^{M-1}\sum\nolimits_{j=0}^{N-1} v_{i,j}\;.
\label{ab_penalty}
\end{equation}
This ensures that even in the worst case when all the items are inadvertently assigned to some knapsack, the penalty values for the QUBO in Eq.~\eqref{eq_qaoa} will still be higher than the negative objective value. These values for the penalty QUBO term work well if all feasible solutions can be obtained by having $H_{\text{single}}=0$ and $H_{\text{capacity}}=0$, \emph{i.e.}, they are hard-constraints as in the standard multi-knapsack QUBO in Eq.~\eqref{eq_qaoa}. However, in the modified QUBO without slack bits, $\tilde{H}_p$ from Eq.~\eqref{eq_alpha_qaoa}, the term $\tilde{H}_{\text{capacity}}$ acts as a soft constraint since there can exist several feasible solutions which do not satisfy $\tilde{H}_{\text{capacity}}=0$ as the knapsacks are not completely filled. Hence, assigning similar coefficients in $H_{\text{single}}$ and $\tilde{H}_{\text{capacity}}$ does not guarantee $H_{\text{single}}$ to be always satisfied. We tested it by optimizing $\tilde{H}_p$ in Eq.~\eqref{eq_alpha_qaoa} using GUROBI's quadratic integer programming solver~\cite{gurobi}. Columns 2-4 of Tab.~\ref{gurobi_1} show the optimal values for the three QUBO terms when using $A$ and $B$ from Eq.~\eqref{ab_penalty}. 
\begin{table}[!htbp]
\begin{center}
\caption{Optimal values for each QUBO term in $\tilde{H}_p$ with $A=B$ as in Eq.~\eqref{ab_penalty}, denoted by $(A)$ and for $A=50\cdot B$, denoted by $(50A)$.}
\begin{tabular}{|C{0.04\textwidth}|C{0.04\textwidth}|C{0.05\textwidth}|C{0.04\textwidth}|C{0.04\textwidth}|C{0.05\textwidth}|C{0.04\textwidth}|}
\hline
\multirow{ 2}{*}{\textbf{Scen.}}&	$\boldsymbol{H_{\text{\textbf{single}}}}$	&	\rule{0pt}{1.3em}$\boldsymbol{\tilde{H}_{\text{\textbf{capacity}}}}$ &	$\boldsymbol{H_{\text{\textbf{obj}}}}$ & $\boldsymbol{H_{\text{\textbf{single}}}}$	&	$\boldsymbol{\tilde{H}_{\text{\textbf{capacity}}}}$ &	$\boldsymbol{H_{\text{\textbf{obj}}}}$\\
& $(A)$ \rule[-1em]{0pt}{0.1em}& $(A)$ & $(A)$ & $(50A)$ & $(50A)$ & $(50A)$\\ \hline 
 0	&	0	&	45	&	-35		&	0	&	45	&	-35	\\
 1	&	0	&	0	&	-2		&	0	&	0	&	-2	\\
 2	&	0	&	0	&	-4		&	0	&	0	&	-4	\\
 3	&	0	&	0	&	-34		&	0	&	0	&	-34	\\
 4	&	0	&	0	&	-30		&	0	&	0	&	-30	\\
 5	&	0	&	0	&	-53		&	0	&	0	&	-53	\\
 6	&	0	&	0	&	-50		&	0	&	0	&	-50	\\
 7	&	0	&	0	&	-51		&	0	&	0	&	-51	\\
 8	&	0	&	0	&	-68		&	0	&	0	&	-68	\\
9	&	0	&	0	&	-71		&	0	&	0	&	-71	\\
10	&	456	&	114	&	-85		&	0	&	4674	&	-53	\\
11	&	472	&	0	&	-89		&	0	&	4012	&	-53	\\
12	&	0	&	320	&	-70		&	0	&	320	&	-70	\\
13	&	0	&	1216	&	-67	&	0	&	1216	&	-67	\\
14	&	0	&	220	&	-45		&	0	&	220	&	-45	\\
15	&	0	&	1968	&	-74	&	0	&	1968	&	-74	\\
16	&	0	&	0	&	-68		&	0	&	0	&	-68	\\
17	&	0	&	0	&	-90		&	0	&	0	&	-90	\\
18	&	0	&	0	&	-105	&	0	&	0	&	-105	\\
19	&	0	&	0	&	-87		&	0	&	0	&	-87	\\ 
\hline
\end{tabular}
\label{gurobi_1}
\end{center}
\end{table}
We can see that equal penalty coefficients lead to infeasibility in $H_{\text{single}}$ qubo terms for some instances. Since $H_{\text{single}}$ models a hard constraint, a higher penalty coefficient is required. Experimenting with different penalty coefficients $A$ and $B$ for $\tilde{H}_p$ we fond that $A = 50 \cdot B$ and $B=\sum_{j=0}^{N-1} w_j + \sum_{i=0}^{M-1}\sum_{j=0}^{N-1} v_{i,j}$ worked the best. We also used the Gurobi solver again with these new penalty coefficients to check if the single penalty term was satisfied. Columns 5-7 of table~\ref{gurobi_1} show the corresponding optimal values for the three QUBO terms when using a factor of $50$ for $A$, \emph{i.e.}, $A=50\cdot B$.

\subsection{Optimizer iterations}\label{sec:res_iterations}
\begin{figure}
\centering
\includegraphics[width=0.45\textwidth]{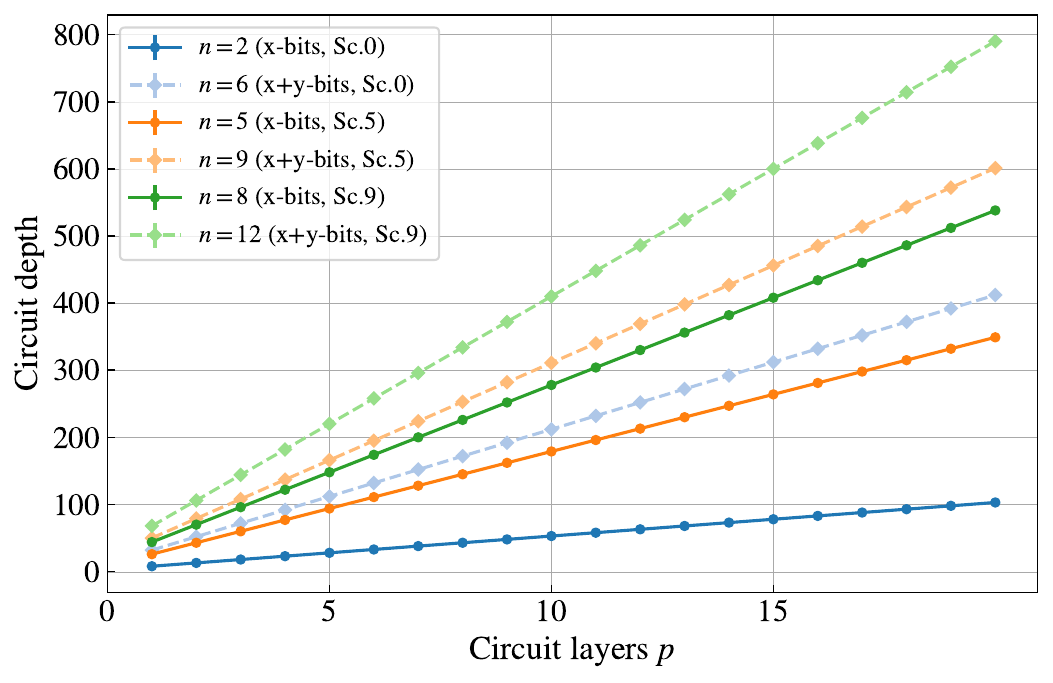}
\caption{Circuit depth when using only x-bits (green circles) and x+y-bits (orange diamonds) vs. scenario number for different number of circuit layers.}
\label{fig:circuit depth}
\end{figure}

In Sec.~\ref{sec:res_algos} we have seen that either the TAE algorithm with slack bits (but without evaluating them in the end) or QAOA without slack bits leads to good results. 
\begin{figure}[htbp]
\centering
\includegraphics[width=0.45\textwidth]{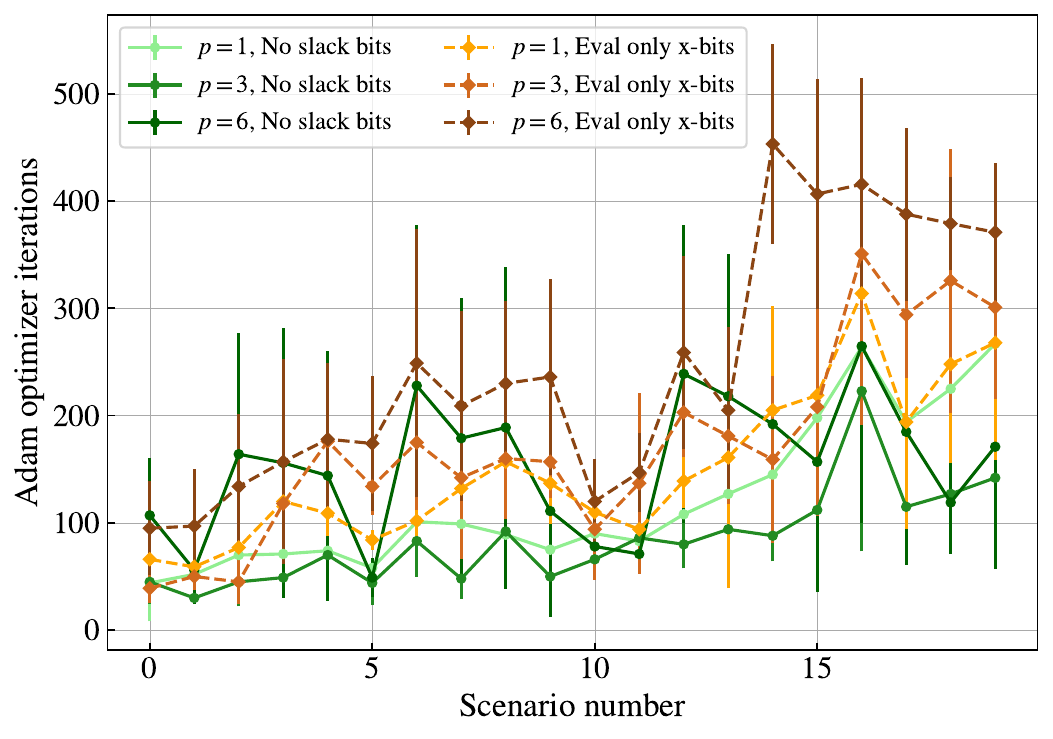}
\caption{Number of adam optimizer iterations for QAOA when using only x-bits (green circles) and x+y-bits (orange diamonds) vs. scenario number. Again, the optimal setting has been used with a single penalty factor $50$ and a sine schedule with $\delta t=0.75$.}
\label{fig:opt_iterations}
\end{figure}
To estimate differences in circuit runtime, we first consider the number of iterations which the Adam optimizer needed until convergence for QAOA, shown in Fig.~\ref{fig:opt_iterations}. In general, the optimization process takes longer for higher number of layers as more parameters need to be optimized. Moreover, discarding the slack bits eases the optimization process, as the circuit becomes simpler and shallower (see also Fig.~\ref{fig:circuit depth}). Even if the slack bits are not evaluated in the end, they are considered in the evaluations during optimization. Depending on the problem size and complexity, the number of optimizer iterations lies in a range of $40-100$ or increases up to $\approx 400$. Moreover, each iteration round in QAOA also includes time for the classical optimizer itself on top of the circuit runtime and measurement time. For TAE, in turn, the circuit only needs to be executed once. 
As discussed in Sec.~\ref{sec:res_algos}, TAE needs more circuit layers to perform equally to QAOA, e.g., TAE with $p=6$ and QAOA with $p=3$ show similar results. Thus, the TAE circuit is longer due to a higher number of layers and a higher number of qubits, including slack bits. In Fig.~\ref{fig:circuit depth}, the circuit depth is plotted vs. number of circuit layers with and without slack bits for scenarios 0, 5, and 9, where $n$ denotes the number of qubits. The circuit depth increases linearly with the number of layers and roughly also with the number of qubits, be it either x- or x+y-bits. Consider scenario 5 as an example: For TAE, slack bits have to be included and the corresponding circuit with $p=6$ has a depth of $195$ (light orange data points). Running QAOA without slack bits and $p=3$ needs a circuit of depth $60$ (dark orange data points), which has to be run $44$ times (green points in Fig.~\ref{fig:opt_iterations}). Assuming that the circuit runtime scales approximately linearly with the circuit depth, we could assign a runtime $T$ to the circuit with $60$ qubits. Then, TAE with slack bits needs about $3.25\,T$, while QAOA without slack bits takes $44\,T$. In general, the overall runtime of QAOA will be much longer due to the classical optimization. However, for near-term devices with only small number of qubits and capable of running only shallow circuits, the repeated execution of a smaller circuit as in QAOA might be beneficial.
\section{Conclusion and Outlook}\label{sec: conclusion and outlook}
In this work we have compared different QUBO variants and objective functions applied to the knapsack problem in two different algorithmic settings, namely QAOA and TAE. 
The standard formulation for the knapsack problem using and evaluating slack variables does not perform well, yielding lower success probabilities than uniform sampling for QAOA and TAE, especially for large scenarios. 
The best results are obtained when completely discarding the slack bits and evaluating the inequality constraint classically, using QAOA: the overall number of variables is smaller, requiring less qubits and thus shallower circuits. Moreover, the fraction of optimal solutions out of all possible bitstrings increases as the size of the state space is decreases.  
Although performing well in many cases, the TAE approach should be used with slack bits in the circuit, since otherwise it can lead to wrong results, as discussed in Sec.~\ref{sec:res_algos}. Here, the variant that only uses x-bits in the final evaluation is recommended, still having the advantage of not relying on convergence of slack bits in the end.

In summary, QAOA without slack bits is better suited for near-term quantum devices, needing less qubits and shallower circuits. The TAE algorithm in turn is expected to run faster than QAOA (see Sec.~\ref{sec:res_iterations}) as no iterations are required, and it is fully quantum, not needing a classical optimization loop. Thus it might be more useful in the future when longer circuits can be executed in a meaningful way. Although the presented extensions of QAOA depict an improvement to the standard algorithm and QUBO formulation, the success probability overall decreases with the problem size and becomes very small already for problems in the range of $20$ qubits, which is still far below any problem size needed in industrial applications. Thus, further improvements of these approaches are necessary. Generlly, the performance increases with more circuit layers, but for TAE we observe no major changes anymore between $p=10$ and $p=20$, indicating a saturation. As shown in~\cite{Hegade_2022}, TAE can be further improved by adding counter-diabatic driving terms to the Hamiltonian. QAOA without slack bits could be combined with other techniques such as constraint preserving mixers~\cite{fuchs_2022}, also with extension to a quantum alternating operator ansatz~\cite{He_2023}, or counter-diabatic driving~\cite{Chandarana_2022}. Moreover, the omission of slack bits and classical evaluation of inequalities could also be used for the Variational Quantum Eigensolver (VQE), since the circuit ansatz for VQE does not require a unitary corresponding to the problem Hamiltonian. While this ansatz does not work for quantum annealing, the idea of only evaluating the x-bits can also be applied there.
 The code which was used to produce the results in this work can be accessed \href{https://github.com/QutacQuantum/effective_inequality_embedding}{on GitHub}.
\section*{Acknowledgement}
We would like to thank Petia Arabadjieva and Davide Vodola for their support and discussions on technical topics during the course of this work.

\bibliographystyle{IEEEtran}
\bibliography{references}


\onecolumn
\appendix

\section{Appendix}

\subsection{Optimizing Annealing Time in TAE}~\label{appendix:opt_annealing_time}
Given a fixed schedule function $s(t)$ and the number of trotter steps $p$, the unitary operator in Equation~\eqref{adb4} depends on the annealing time $\mathcal{T}$. In this work we also implement the variational version of TAE, where in, we optimize $\mathcal{T}$. This variational TAE has also been studied by He \emph{et al.} in~\cite{He_2023}. This approach helps in maintaining the variational aspect of QAOA and comprises of only a single variational parameter to be optimized by the classical subroutine.

\subsection{Comparison of annealing schedules}~\label{appendix:annealing_schedules}

\begin{figure}[H]
\centering
\includegraphics[width=1\textwidth]{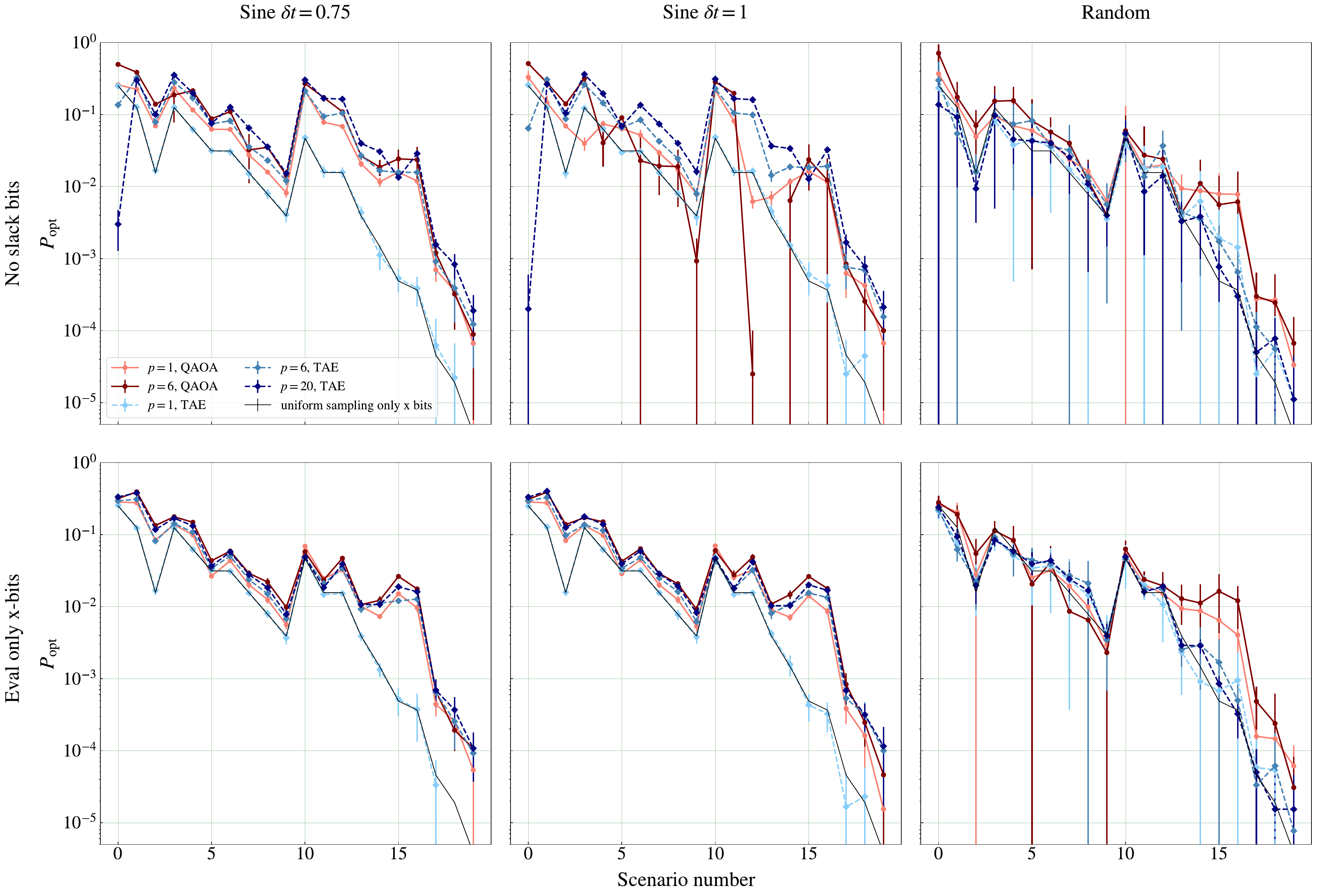}
\caption{Total probability of sampling optimal solutions $P_{\mathrm{opt}}$ for different objective functions (upper and lower panels) vs. scenario number. The columns indicate different annealing schedules used for TAE and as initial parameters for QAOA. The results for QAOA (red circles) and TAE (blue diamonds) for various numbers of circuit layers are shown. As a baseline, $P_{\mathrm{opt}}$ for randomly drawing x-bitstrings is plotted as a solid black line. Missing points are zero and thus not displayed in the logarithmic scale.}
\label{fig:compare_schedules}
\end{figure}

We tested different annealing schedules, being used directly in TAE or as initial parameters for QAOA. In Fig.~\ref{fig:compare_schedules}, the probability of sampling the optimal solution is plotted against the scenario number for QAOA and TAE comparing a sinusoidal schedule with $\delta t = 0.75$ (first column), a sinusoidal schedule with $\delta t = 1$ (second column) and random angles (third column). The first row contains results for discarding the slack bits, the second row shows the results for using the slack bits in the circuit but not evaluating them. We can clearly see that the sinusoidal schedules perform much better than random angles, not only for TAE, but also for QAOA. Changing the step-size only influences the results without slack bits, where especially QAOA performs better for $\delta t = 0.75$.

\pagebreak

 \subsection{Knapsack instances}\label{appendix: knapsack instances}
 \begin{table*}[htbp]
     \centering
         \caption{Overview of knapsack instances}
     \begin{tabular}{|c ||c|c|c|c|c|c|c|c|c|}
     \hline
         \textbf{Scen. no.} & Knapsacks & Capacity & Items  & Item values & Item weights & Log. bits & Slack bits & Opt. val. & No. opt.\\ \hline \hline
         0 & 1 & (9) & 2 & (19, 16) & (4, 6)& 2 & 4 & 19 & 1\\ \hline 
         1 & 1 & (3) & 4 & (4, 4, 1, 2) & (2, 2, 2, 3) & 4 & 2 & 4 & 2 \\ \hline
         2 & 1 & (3) & 6 & (1, 3, 5, 2, 4, 1) & (3, 2, 2, 2, 3, 2) & 6 & 2 & 5 & 1 \\ \hline
         3 & 1 & (10) & 4 & (19, 19, 17, 17) & (6, 6, 3, 7) & 4 & 4 & 36 & 2 \\ \hline
         4 & 1 & (8) & 5 & (15, 15, 16, 17, 17) & (2, 6, 5, 5, 4) & 5 & 4 & 32 & 2 \\ \hline
         5 & 1 & (8) & 5 & (18, 17, 19, 18, 19) & (2, 4, 5, 2, 3) & 5 & 4 & 55 & 1 \\ \hline
         6 & 1 & (10) & 6 & (19, 17, 15, 17, 15, 18) & (7, 1, 6, 7, 5, 3) & 6 & 4 & 50 & 2 \\ \hline
         7 & 1 & (8) & 6 & (19, 18, 16, 18, 17, 16) & (6, 7, 4, 3, 3, 2) & 6 & 4 & 51 & 1 \\ \hline
         8 & 1 & (8) & 8 & (17, 16, 17, 15, 18, 17, 16, 18) & (5, 4, 1, 5, 4, 1, 2, 3) & 8 & 4 & 68 & 2 \\ \hline
         9 & 1 & (9) & 8 & (16, 17, 17, 19, 18, 16, 17, 19) & (3, 2, 5, 1, 4, 4, 1, 4) & 8 & 4 & 72 & 1 \\ \hline
         10 & 2 & (11, 8) & 3 & [(19, 16, 16); (19, 16, 18)] & (2, 4, 4) & 6 & 8 & 53 & 3 \\ \hline
         11 & 2 & (8, 11) & 3 & [(18, 19, 17); (18, 17, 18)] & (3, 6, 2) & 6 & 8 & 55 & 1 \\ \hline
         12 & 2 & (9, 9) & 4 & \begin{tabular}{@{}c@{}}[(19, 16, 19, 16); \\ (19, 16, 19, 16)]\end{tabular}  & (4, 6, 4, 6) & 8 & 8 & 54 & 4 \\ \hline
         13 & 2 & (10, 10) & 4 & \begin{tabular}{@{}c@{}}[(18, 16, 15, 19); \\ (16, 17, 15, 16)]\end{tabular}  & (7, 1, 5, 7) & 8 & 8 & 52 & 1 \\ \hline
         14 & 2 & (8, 8) & 6 & \begin{tabular}{@{}c@{}}[(15, 15, 18, 15, 15, 18); \\ (15, 15, 18, 15, 15, 18)]\end{tabular} & (7, 4, 3, 7, 4, 3) & 12 & 8 & 66 & 6 \\ \hline
         15 & 2 & (8, 8) & 6 & \begin{tabular}{@{}c@{}}[(19, 17, 18, 19, 17, 18); \\ (19, 17, 18, 19, 17, 18)]\end{tabular} & (5, 5, 5, 5, 5, 5) & 12 & 8 & 38 & 2 \\ \hline
         16 & 2 & (10, 10) & 8 & \begin{tabular}{@{}c@{}}[(19, 19, 17, 17, 19, 19, 17, 17); \\ (19, 19, 17, 17, 19, 19, 17, 17)]\end{tabular} & (6, 6, 3, 7, 6, 6, 3, 7) & 16 & 8 & 72 & 24 \\ \hline
         17 & 2 & (11, 9) & 8 & \begin{tabular}{@{}c@{}}[(19, 19, 17, 15, 19, 15, 16, 15); \\ (19, 15, 18, 15, 15, 18, 18, 17)]\end{tabular} & (1, 6, 4, 5, 7, 3, 4, 5) & 16 & 8 & 91 & 3 \\ \hline
         18 & 2 & (9, 10) & 9 & \begin{tabular}{@{}c@{}c@{}}[(15, 18, 15, 17, 16, 18, \\16, 16, 15);  (18, 15, 18,\\ 17, 16, 16, 19, 18, 17)]\end{tabular} & (3, 5, 2, 3, 2, 6, 5, 2, 7) & 18 & 8 & 105 & 5 \\ \hline
         19 & 2 & (8, 9) & 9 & \begin{tabular}{@{}c@{}c@{}}[(18, 15, 15, 16, 17, 16, \\ 19, 15, 17); (18, 15, 16, \\ 16, 16, 18, 15, 19, 15)]\end{tabular} & (1, 5, 1, 1, 5, 7, 7, 5, 3) & 18 & 8 & 103 & 1 \\ \hline
         20 & 3 & (9, 9, 9) & 6 & \begin{tabular}{@{}c@{}c@{}}[(19, 16, 19, 16, 19, 16); \\ (19, 16, 19, 16, 19, 16) \\(19, 16, 19, 16, 19, 16)]\end{tabular} & (4, 6, 4, 6, 4, 6) & 18 & 12 & 73 & 54 \\ \hline
         21 & 3 & (9, 10, 11) & 6 & \begin{tabular}{@{}c@{}c@{}}[(19, 16, 19, 17, 17, 19); \\ (16, 17, 19, 17, 18, 16) \\(15, 16, 19, 17, 17, 19)]\end{tabular} & (7, 6, 7, 
         5, 5, 4) & 18 & 12 & 92 & 1 \\ \hline
     \end{tabular}
     \label{tab:my_label}
 \end{table*}

\end{document}